\documentstyle[12pt]{article}

\def\abstracts#1#2#3{{
        \centering{\begin{minipage}{5in}\baselineskip=12pt\tenrm
        \centerline{ABSTRACT}
        \parindent=0pt #1\par
        \parindent=15pt #2\par
        \parindent=15pt #3
        \end{minipage} }\par}}

\normalsize
\def\sp{~~~~~}
\def\a{\alpha}
\def\b{\beta}

\def\d{\delta}
\def\e{\epsilon}
\def\f{\phi}
\def\g{\gamma}
\def\h{\eta}

\def\j{\psi}
\def\k{\kappa}
\def\l{\lambda}
\def\m{\mu}
\def\n{\nu}
\def\o{\omega}
\def\p{\pi}

\def\r{\rho}

\def\t{\tau}

\def\D{\Delta}

\def\G{\Gamma}

\def\L{\Lambda}
\def\O{\Omega}

\def\Q{\Theta}
\def\S{\Sigma}

\def\cl{{\cal L}}
\def\cm{{\cal M}}

\def\co{{\cal O}}

\def\rt{\rightarrow}

\def\pa{\partial}

\def\bar#1{\overline{#1}}
\def\Tilde#1{\widetilde {#1}}
\def\Hat#1{\rlap{\kern.10em$\widehat{\phantom G}$}#1}
\def\HAt#1{\rlap{\kern.05em$\widehat{\phantom G}$}#1}

\def\cap#1{\rlap{\kern.1em$\widehat{\phantom{G\vrule height.8em}}$}#1{}}
\def\Cap#1{\rlap{\kern.05em$\widehat{\phantom{G\vrule height.8em}}$}#1{}}

\newcounter{sxn}

\newcounter{axn}

\def\br{\begin{thebibliography}{abc}}
\def\er{\end{thebibliography}}

\date{}

\begin{document}

\bibliographystyle{unsrt}
\thispagestyle{empty}
\normalsize
\begin{flushright}
SU-4240-523\\
November 1992
\end{flushright}
\vglue 1.0cm
\centerline{ \tenbf SYMMETRY BREAKING IN A GENERALIZED SKYRME
MODEL\footnote{Talk at Workshop on ``Baryons as Skyrme Solitons'',
September 28-30, 1992; Siegen, Germany.}}
\vglue 1.0cm
\begin{center}
{ \tenrm J. SCHECHTER }
\end{center}
\vglue 0.5cm
\begin{center}
{\it Department of Physics, Syracuse University,\\ Syracuse,
NY 13244-1130,USA\\}
\end{center}
\vglue 1.0cm

\abstracts{
We first outline the calculations of the neutron-proton mass difference and of
the axial singlet matrix element (relevant to the ``proton spin" puzzle) in a
generalized Skyrme model of pseudoscalars and vectors.  These two calculations
are, perhaps surprisingly, linked to each other and furthermore are sensitive
to some fine details of symmetry breaking in the effective meson Lagrangian.
This provides a motivation for us to examine these symmetry breaking terms more
closely.  We find a qualitatively new feature in the symmetry breaking pattern
of the vector meson system and discuss its significance.}{}{}
\vglue 1.0cm
{\bf\noindent 1. Introduction}
\vglue 0.5cm
\baselineskip=14pt
\rm
This material is based on work with Anand Subbaraman and Herbert Weigel which
will be described in more detail elsewhere$^1$  The original motivation
was to update some older papers$^{2,3,4}$, which should be consulted for
adequate references to background work.

There are three reasons for taking the Skyrme model approach seriously. a) It
{\em works} fairly well for baryon mass {\em differences}$^5$, static
properties
   ,
scattering amplitudes etc. b) the picture is suggested by the consistent
large $N_c$ approximation scheme for QCD. c) It makes use of a nice feature
of ``old physics''-- the ``pion cloud'' of the Yukawa theory -- and builds upon
it in a constructive way.  Roughly speaking, the original Skyrme model
replaces the ``core'' of the nucleon by a boundary condition.  We shall
consider here that the vector mesons describe more of the core physics.  Other
approaches utilize explicit quarks for this purpose.

The calculation of the non-electromagnetic part of the neutron-proton mass
difference (as given in ref. 2) is reviewed in section 2.  This has the
interesting aspect that it forces one to go beyond the Skyrme model of pions to
include at least the $\h$ field and also some information (via the introduction
of  vector mesons, for example) about ``short distance'' physics in the model.
The extension to the full SU(3) model of pseudoscalars and vectors is, in fact,
indicated.

In section 3, the calculation $^{3,6}$ of the protons's axial singlet
matrix element in the pseudoscalar-vector model is briefly reviewed and the
connection with the so-called ``proton-spin puzzle$^7$'' is given.  It
is shown how a conjectured decomposition of the matrix element into ``matter''
and ``glue'' parts is related$^4$ to the  calculation of the $n-p$
mass difference in this model.

In order to improve the accuracy of the calculations discussed in sections 2
and 3, or at the least, to test their sensitivity to reasonable parameter
changes it is necessary to employ the {\em full three flavor}
pseudoscalar-vector
model.  The first step of carefully discussing symmetry breaking in the
underlying meson Lagrangian is given in section 4.  It is noted that the
introduction of derivative type symmetry breaking terms for the vectors
dramatically improves the agreement with experiment.  This encourages us to
suggest that the three-flavor chiral perturbation theory program$^8$
be extended to include the vectors as well as pseudoscalars so that it has a
chance of describing low energy physics up to about 1 GeV.  The relevant
soliton calculations, using the full three flavor meson Lagrangian, are
discussed in the following talk of Herbert Weigel$^9$.
\vglue 1.0cm
{\bf\noindent 2. Neutron-proton mass difference}
\vglue 0.5cm

We write
$$m(n)-m(p)=\D_{EM}+\D \eqno(2.1)
$$
\noindent
where $\D_{EM}$ is due to photon exchange and $\D$ is due to the different $u$
and $d$
quark masses in the fundamental QCD Lagrangian.  Subtracting the conventional
estimate$^{10}$ for  $\D_{EM}$ gives $\D=(2.05 \pm 0.30)$ MeV as the number
to be understood in
the present calculation.  Let us first try to explain this at the two flavor
level, which seems very reasonable.  But there is a problem:  In the original
Skyrme model containing only pions, the mesonic term mocking up the fundamental
iso-spin breaker has the form ${\rm Tr}[\t_3(U+U^\dagger)]$ which vanishes
identically when we plug in
$U=\cos \o+i{\underline n}\cdot{\underline \t}\sin\o$ as the
general parameterization of a $2 \times 2$ unitary unimodular matrix.  A
way out at the two flavor level would appear to be to include also an isoscalar
meson, $\h\sim(\bar{u}u+\bar{d}d)/ \sqrt{2}$. In this case $U$ is no longer
unimodular and  the mesonic symmetry
breaker does not vanish when we plug in the more general form
$U=e^{i\chi} (\cos \o +i{\underline n}\cdot {\underline \t \sin \o})$.
In particle language we get a symmetry breaker like $\e\h\p^0$.  We would
expect
    an $n-p$
splitting in the Yukawa theory from graphs in which the nucleon emits a virtual
$\p^0$ which converts, due to the $\e\h\p^0$ interaction, into an $\h$ which is
then reabsorbed by the nucleon.
Unfortunately the nucleon matrix elements of the operator $\e\h\p^0$
 vanish for the
original Skyrme Lagrangian; the $\h$ field does not get excited.
This is because the $\h$ field only appears
quadratically in the Lagrangian and, for example, the static
Hamiltonian terms
$\frac{1}{2}(\bigtriangledown\h)^2+\frac{1}{2}m^2_\h\h^2$ are minimized for
$\h=
   0$.  Clearly a term
linear in $\h$ is required to act as a source (the term $\e\h\p^0$ does so
to a small
extent but leads to $\D=\co(\e^2)$ which is negligible).
It is interesting that exciting an $\h$ forces us to extend our original model.
We may think of this effect as arising from the ``core" of the nucleon,
described, for example, by vector mesons.  It was shown that in a model of
pseudoscalars {\em and} vectors there are terms$^{11}$ linear in
$\h$ proportional to the Levi-Civita tensor,
$$\e_{\m\n\a\b} \pa_\m\h(\o_\n\pa_\a \o_\b+~{\rm many ~others})~.$$
\noindent
These terms do excite the $\h$ field to the required strong order when account
is taken of the need to ``crank'' the $\h$ field.  This amounts to allowing
the $\h$ to
get excited by centrifugal effects; the ansatz
$$\h({\underline r})=\h(r){\underline {\hat r}} \cdot {\underline\O},\eqno(2.2)
$$
\noindent
where $\frac {i}{2} \t_a\O_a=A^\dagger \dot{A}$ is the angular velocity,
($A$ is defined from
$U(t)=A(t)U_c({\underline x})A^\dagger(t)$, $U_c(x)$
being the classical soliton solution) is substituted in
and the profile $\h(r)$ is chosen so as to maximize the moment of inertia.
Using
(2.2) in the symmetry breaking terms finally gives an iso-spin breaking piece
in the collective Hamiltonian of the form
$$H_{SB} = - \D I_3, \eqno (2.3)
$$
\noindent
where $I_3$ is the iso-spin {\em operator}.  Note that there will now be a
number of
different types of symmetry breaking terms in our effective meson Lagrangian.
For example there will also be an $\o\r^0$ transition piece and others
involving
more derivatives.  Several years ago, $\D$ was estimated$^2$ on this
basis to be about 1.3 MeV, just a little low compared to (2.1).  It was also
shown that the extra amount was likely due to inclusion of the strange degree
of freedom.  This is plausible in the Yukawa theory since $K$ meson exchange
with unequal $K^+$ and $K^0$ meson masses is also expected to contribute to
 $\D$.

However there are a number of interesting complications when we consider the
three flavor case in the generalized Skyrme model approach.  First, of course,
we have both $\h$ and $\h^\prime$ fields appearing.  Structurally, the
$\h^\prime$ appears in the
mesonic Lagrangian as the phase of the $3 \times 3$ matrix $U$ in analogy
to the $\h$ in
the two flavor case.  But there is now a qualitatively new feature in that one
has a contribution to $\D$ even without cranking  the $\h$.
 This is proportioned to
$$
<n|D_{38}(A)|n>-<p|D_{38}(A)|p>, \eqno (2.4)
$$

\noindent
where $D_{ab}(A)=\frac{1}{2}~{\rm Tr}~(\l_a A\l_b  A^\dagger)$  is the
SU(3) adjoint representation matrix element.  The $\l_a$ are the
Gell-Mann matrices.  Evaluating (2.4) using for $|n>$ and $|p>$ the eigenstates
   of
the symmetric  (no symmetry breaking terms) collective Hamiltonian of the
pseudoscalar only Skyrme model actually gives the fairly large value of about
1.3 MeV as the contribution to $\D$.  This might suggest that perhaps there was
no need to excite the $\h$ field by extending the model to include vectors.
However that is wrong.  Yabu and Ando$^{12}$ showed that the baryon
eigenstates are very sensitive to symmetry breaking. When the collective
Hamiltonian is diagonalized {\em exactly} the contribution to $\D$ is reduced
by
a factor of $\frac{1}{2}$. Furthermore, improving the Yabu-Ando calculation
by cranking the
$K$ mesons leads$^{13}$ to another factor of $\frac{1}{2}$ reduction (in the
mod
   el of pseudoscalars
only).  Altogether we thus {\em estimate}
$$
\D\approx 1.3 + (1.3)/4\approx 1.6~{\rm MeV}~, \eqno(2.5)
$$

\noindent
where the first piece is due to cranking the $\h$ at the two flavor level and
 the
second piece is due to (2.4).  A similar estimate was presented in ref. 2.  We
stress that this is just an estimate since the first part was calculated in a
two flavor model of vectors and pseudoscalars while the second was gotten from
a three flavor model of pseudoscalars only.  It is thus desirable to see what
the result would be in a treatment using the full three flavor model with both
pseudoscalars and vectors throughout.  This is part of our motivation  for a
more through analysis of symmetry breaking in the full model.
\vglue 1.0cm
{\bf\noindent 3. Axial Singlet Matrix Element}
\vglue 0.5cm
\baselineskip=14pt
\rm
The current interest in this ``hot topic''  will turn out to perhaps justify
the complicated meanderings to which we have exposed the reader in the last
section.  The form factors of the $a$th flavor axial current in the proton are
defined by
$$\sqrt{\frac{p_0p_0^\prime V^2}{M^2_N}} <p'|i \bar q_a \g_\m\g_5q_a|p>
=i\bar {u} (\underline {p}^\prime) [\g_\m\g_5H_a(q^2)+\frac {iq_\m}{2M_N} \g_5
\Tilde {H}_a(q^2)]u(\underline {p}).\eqno (3.1)
$$

The {\em singlet} form factors are defined by
$$
H(q^2)=\sum_{a=1}^{3} H_a (q^2);~\Tilde{H}
(q^2)=\sum_{a=1}^{3}\Tilde{H}_a(q^2),~\eqno(3.2)
$$

\noindent where $(1,2,3)=(u,d,s)$.

A great deal of excitement was caused by the EMC (European Muon Collaboration)
experiment$^{14}$ on deep inelastic polarized $\m$ - polarized $p$ scattering.
The result can be interpreted as a measurement of
$$
\frac{4}{9} H_1(0) + \frac{1}{9} H_2(0) +
\frac {1}{9} H_3(0).\eqno(3.3)
$$

\noindent Another linear combination of these form factors is known from
neutron beta decay:
$$
H_1(0)-H_2(0)\simeq 1.25, \eqno (3.4)
$$

\noindent
while a third can be estimated from hyperon beta decay plus SU(3) flavor
covariance:
$$
R\equiv H_1(0)+H_2(0)-2H_3(0)\approx 0.68. \eqno(3.5)
$$

\noindent For the first time it was possible to get information on the matrix
element of the axial {\em singlet} current, $J^5_\m$.  This is especially
interesting since $J^5_\m$ obeys the {\em anomalous} divergence equation
$$
\pa_\m J^5_\m=2i\sum_{a=1}^{3} m_a \bar{q}_a\g_5  q_a + \pa_\m K_\m,
\eqno(3.6)$$

\noindent where $K_\m$ is the Chern-Simons current of the Yang-Mills theory.
Note that $\pa_\m J^5_\m$ does not vanish when the $m_a$ go to zero.  From the
EMC result for (3.3) together with (3.4) and (3.5) one gets:
$$
H(0)  =  0.03 \pm 0.18, \eqno(3.7a)
$$
$$
H_3(0) = -0.22 \pm 0.06. \eqno (3.7b)
$$

\noindent Eq. (3.7a) is surprising since it is expected to be 1.0 in the naive
non-relativistic quark model wherein $\frac{1}{2} H(0)$ is identified as the
proton expectation value of the {\em quark spin} part of the total angular
momentum operator.  Also, (3.7b) is surprising since the ``strangeness
content'' of the proton is expected to be  small.  It seems to us that, of
these two surprising results, only the first is a reliable deduction  from the
measurement of (3.3).  That is because the use of flavor SU(3) to obtain
$R$
in (3.5) is debatable.  The SU(3) Skyrme model, in fact, suggests$^{15}$ much
smaller values of $R$ (while not giving big SU(3) deviations for the flavor
{\em changing} currents which enter into the Cabibbo theory).  Taking, somewhat
arbitrarily, $R=0.3$ would lead to the deduction from experiment
$$
H(0) = 0.12 \pm 0.18, \eqno(3.8a)
$$
$$
H_3(0) = -0.06 \pm 0.06. \eqno(3.8b)
$$
\noindent Eq. (3.8a) is {\em still} small and surprising while (3.8b) is no
longer surprising!

Now in the Skyrme model containing just pseudoscalar fields, there is a nice
result$^{16}$:
$$
H(0)=0.\eqno (3.9)
$$

\noindent So, what is {\em mysterious} in the quark model is the {\em expected}
 feature in the Skyrme model.  Let us try to understand this from a
different$^{17}$ point of view.  In the Skyrme model of psudoscalars the $\h'$
field enters as $U=\exp \left (\frac {2i\h'}{\sqrt{3}F_\p}\right ) \Tilde {U}$,
where $\det \Tilde U=1$.  By Noether's theorem the axial singlet current is
simply
$$
J^5_\m=-\sqrt{3}F_\p \frac{\pa \cl}{\pa(\pa_{\m}\h')}=\sqrt{3} F_\p\pa_{\m}\h',
\eqno(3.10)
$$

\noindent where the second step holds because $\h'$ enters only quadratically
in $\cl$ (we just pick up its kinetic term for (3.10)).  Now the result (3.9)
is
evident since $J^5_\m$ is a pure gradient and hence proportional to $q_\m$ in
momentum space.  From the form factor decomposition in (3.1) we see that (3.10)
only contributes to the ``induced'' form factor $\Tilde H (q^2)$ and hence
$H(q^2)$ must vanish in this simple model.

But this is troublesome from the point of view of the $n-p$  calculation
described in the last section.  There we {\em needed} a term
linear in $\h'$ in order to excite the pseudoscalar iso-singlet.
We saw that there {\em was} such a term when vectors were added; it can be
written as
$$
\cl=... - \frac {1}{\sqrt{3}F_{\p}} \pa_\m \h^\prime \Tilde{J}^5_\m.
\eqno(3.11)
$$

\noindent Here $\Tilde{J}^5_\m$ is proportional to $\e_{\m\n\a\b}$ and is due
to the vector fields.  It can be identified, using (3.10), as a new (short
distance) contribution to $J^5_\m$:
$$
J^5_\m=\sqrt{3} F_\p\pa_\m\h^\prime + \Tilde {J}^5_\m. \eqno(3.12)
$$

\noindent Since $\Tilde J^5_\m$ is not a pure gradient$^{18}$ it does
contribute to $H(0)$.  Does this destroy  the nice prediction of the Skyrme
model?  A couple of years ago it was found that $H(0)\approx 0.30$ in this
model.$^{3,6}$  This is still qualitatively small (and in agreement with
experiment if (3.8a) is taken rather than (3.7a)).

Another aspect of this problem which the Skyrme model approach might illuminate
concerns the so-called ``2-component decomposition''.$^{19}$  This is an
attempt to make the small value of $H(0)$ intuitively plausible from the QCD
parton model point of view.  It is certainly legal to write:
$$<p|J^5_\m|p>=<p|(J^5_\m-K_\m)|p>+<p|K_\m|p>.\eqno(3.13)
$$

\noindent Since the first term on the right hand side is conserved when the
light quark masses are neglected it is tempting to consider it, in some sense,
the usual ``matter'' contribution and to be around the naive quark model value
of 1.  It is hoped that this would be largely cancelled by the second ``glue''
term.  There has been considerable discussion of this possibility in the
literature.$^{20}$  One criticism is that the decomposition (3.13) is not gauge
invariant.  It has been suggested that a gauge invariant and hence better way
is to give an analogous decomposition based on a generalized Goldberger-Trieman
relation.  This reads$^{21}$:
$$
H(0)=\frac{\sqrt{3}F_\p}{2M_N}\left (g_{\h^{\prime}NN}-\sqrt{3}F_\pi
m^2_{\h^{\prime}}g_{GNN}\right ), \eqno(3.14)
$$

\noindent where $g_{\h^{\prime}NN}$ is the Yukawa coupling constant of the
(pure SU(3) singlet) $\h^{\prime}$ field with the nucleons while $g_{GNN}$ is
the
Yukawa coupling constant for the composite glue field $G\equiv  \pa_\m K_\m$.
The first term is supposed to be the ``matter'' contribution, the second the
``glue'' piece.  Now it is very difficult$^{22}$ to obtain a reliable value for
$g_{\h^{\prime}NN}$ from experiment and even harder to get an experimental
handle on $g_{GNN}$.  However it is possible to make a theoretical estimate in
the generalized Skyrme model of pseudovectors and vectors.  We showed$^{4}$
that the existence of the second term in (3.14) corresponds to the presence
of a new term,
$$
\frac {t}{3F^{2}_{\p}m^{2}_{\h^{\prime}}} \pa_\m G \Tilde{J}^5_\m, \eqno(3.15)
$$

\noindent in the effective Lagrangian. $t$ is a new dimensionless constant.
The effect of this term is to provide an extra contribution to the
$\h^{\prime}$ excitation (the ``auxiliary'' field G must be eliminated by its
equation of motion in terms of $\h^{\prime}$ to see this) but {\em not} to the
current, $J^5_\m$.  This means that the $\h$ contribution to $\D$ in
(2.3) should be multiplied by a factor of $(1-t)$.   Thus we can try to improve
the predicted value of $\D$ by adjusting $t$.  This predicts the coupling
constants in the two component decomposition as$^4$:
$$
g_{\h^{\prime}NN} = (1-t) \frac{2M_{N}H(0)}{\sqrt{3}F_\p},
$$
$$
g_{GNN} =
\frac{t}{t-1}\frac{g_{\h^{\prime}NN}}{\sqrt{3}F_\p m_{\h^{\prime}}}.\eqno(3.16)
$$

\noindent Here $H(0)$ should still be taken to be $0.30$ as discussed after
(3.12).  Finally, ``improving'' the prediction for $\D$ (assuming the central
value, $2.05$ MeV) gives$^4$ for (3.14):
$$
H(0)=~\rm{``matter''}~ + ~{\rm``glue''}
$$
$$
0.3=0.4 - 0.1. \eqno(3.17)
$$

\noindent This indicates a rather small gluonic as well as a small matter
contribution.  We should stress, that since this estimate is related to perhaps
arcane details of symmetry breaking and $\h^\prime$ excitation, (see the
discussion at the end of section 2) it is desirable to investigate further the
sensitivity of the result to these factors.
\vglue 1.0cm
{\bf\noindent 4. Back to the Meson Lagrangian}
\vglue 0.5cm

The first step in a more detailed investigation of symmetry breaking in the
generalized Skyrme model is clearly to examine the underlying meson Lagrangian
itself. Since the soliton energy is in the neighborhood of 1 GeV it would seem
prudent for the meson Lagrangian to adequately describe mesonic physics up to
this energy. This would imply that we include in addition to the pseudoscalars,
the vectors (because, among other things, they are ``there'').  We will also
include the composite glue field $G$ in order to to simplify the implementation
of the $U(1)$ axial anomaly. (A composite scalar glue field, $H=\Q_{\m\m}$ can
also, if desired, be introduced in order to conveniently implement the trace
anomaly).

The dynamical variables we use in the Lagrangian are, in addition to the chiral
field $U$, $\xi\equiv U^{1/2}$ and a vector meson nonet, $\r_\m$ related to
linearly transforming objects $A^L_\m$ and $A^{R}_\m$ by
$$
A^L_\m= \xi \r_\m\xi^\dagger+\frac{i}{\tilde{g}} \xi
\pa_\m\xi^\dagger;~A^{R}_\m=\xi^{\dagger}\r_\m\xi+\frac{i}{\tilde
{g}}\xi^\dagger \pa_\m \xi.\eqno(4.1)
$$
The chiral symmetric and OZI rule conserving terms of the Lagrangian have been
given in many places.$^{23}$ Here we shall concentrate our discussion on the
symmetry breaking terms.  For this purpose we want to introduce some notation
associated with the mass terms of the fundamental QCD Lagrangian:
$$
\cl_{\rm mass}= -m_u \bar {u}u-m_d\bar{d}d-m_s\bar {s}s \equiv -
\hat{m}\bar{q}\cm q, \eqno(4.2)
$$
\noindent where $\hat{m}= (m_u+m_d)/2$ and the dimensionless matrix $\cm$
is written as
$$\cm=y \l_3 + T + xS,$$
$$\l_3=~{\rm diag}~ (1,-1,0), ~T=~{\rm diag}~(1,1,0),~S=~{\rm diag}~(0,0,1),$$
$$x= m_s/\hat{m},~ y= \frac {m_u-m_d}{m_u+m_d}. \eqno (4.3)
$$

\noindent
In addition we note the redundant, but convenient, quantity$^{10}$,
$$
R = \frac{m_s-\hat{m}}{m_d-m_u} = \frac{{\rm ``strange"~splitting}}
{{\rm iso-spin~ splitting}}. \eqno(4.4)
$$

\noindent The interest of $R$ lies in the fact that it may, in principle, be
straightforwardly extracted from SU(3) multiplets other than $0^-$ and $1^-$.
We have introduced so much notation for $\cm$ since all  symmetry breaking
terms will be taken to be proportional to it.  Rather than writing down
all symmetry breaking terms as in the CPT (chiral perturbation theory)
program$^8$ of pseudoscalars only, we shall only write the ones we believe
are most important. These are the ones which (except for those needed to
solve the $\h^{\prime}$ problem ) obey the OZI rule. Incidentally, we may
note that the OZI violating symmetry breaking terms turn out to have small
coefficients in the CPT approach too. We implement the OZI rule in Okubo's
original form$^{24}$: all mesons are described by nonets and only terms which
are a single trace in flavor space are included.  This eliminates the
``hairpin'' diagrams. Without further ado we list the OZI rule conserving,
symmetry breaking terms to be included in the effective Lagrangian:
$$
\a^\prime ~{\rm Tr}~[\cm(A^L_\m UA^R_\m+A^R_\m U^\dagger A^L_\m)]
+  \b^\prime~{\rm Tr}~[\cm(\pa_\m U\pa_\m U^\dagger U+U^\dagger
\pa_\m U \pa_\m U^\dagger)]$$
$$ + \g^\prime ~{\rm Tr}~[\cm(F^L_{\m\n}UF^R_{\m\n}+F^R_{\m\n}U^\dagger
F^L_{\m\n})]
+  \d^\prime ~{\rm Tr}~ [\cm(U+U^\dagger - 2)]$$
$$+ \l^{\prime^{2}} ~{\rm Tr}~ [\cm U^\dagger \cm U^\dagger+\cm U \cm
U-2\cm^2], \eqno(4.5)
$$

\noindent where
$$F^{L,R}_{\m\n}=\pa_\m A^{L,R}_\n -
\pa_\n A^{L,R}_\m-i\tilde{g}[A^{L,R}_\m,A^{L,R}_\n].$$

\noindent The explanation for these terms is as follows. First, the $\d'$
term is the
usual, main symmetry breaker for the pseudoscalars. The $\a'$ term plays the
same role for the vectors.
However, noting (4.1), the $\a^\prime$ term also supplies an undesirably
large amount of derivative-type symmetry breaking for the pseudoscalars. Most
of this is cancelled by the $\b^\prime$ term. Together, the $\a^\prime$
and $\b^\prime$ terms enable us to fit both $F_K$ and $F_\p$. The
$\g^\prime$ term is a derivative type symmetry breaker for the vectors. It
was set to zero for simplicity in ref. 2 but we shall see here that it may
be rather important. Finally, the $(\l^\prime)^2$ term, though of second
order in $\cm$, is argued to be of the same order as the $\b^\prime$ term
in the CPT program$^8$ ($\cm$ counts as two derivatives). This is reasonable
because the symmetry breaking for the pseudoscalars is relatively large. Such
a term may be less important for the vectors but, in any event, it would have
the form $\m^\prime ~{\rm Tr}~ (\cm A^L_\m \cm A^R_\m)$.

Now our task is to fit the quantities
$$
\a^\prime ,\b^\prime , \g^\prime ,\d^\prime ,\l^{\prime 2};
{}~\tilde{g},m_v;~x=\frac {m_s}{\hat{m}},~y=\frac {m_u-m_d}{m_u+m_d},
$$

\noindent where $\tilde{g}$ is the ``gauge coupling constant'' in (4.1)
and $m_v$ is a ``bare'' vector meson mass, from the experimental mass spectrum
of $0^-$ and $1^-$ mesons, the $0^-$ decay constants and the $V \rt \f\f$ decay
widths.  Since there are many unknowns one might think that the work could be
simplified if we had reliable information on the quark mass ratios $x$ and $y$
from outside the $0^-$ and $1^-$ multiplets.  Let us make a small {\em detour}
to see if this is possible.

Certainly the best source of information is the baryon octet.  One may make a
perturbation around zero quark masses, $m_u=m_d=m_s=0$ (chiral perturbation
theory) or around, say, the symmetric point
$$m_u=m_d=m_s=\frac{1}{3}(m_u+m_d+m_s), \eqno (4.6)
$$
which we may denote as ``flavor'' perturbation theory.  We can then use group
theory (generalized Wigner-Eckart theorem) to derive mass relations in a well
known way.  To first order in perturbation it doesn't matter which scheme is
used and we find the Gell-Mann Okubo formula
$$
\L-N=\frac{1}{3} [2(\Xi-N)-(\S-N)]
$$
$$176.9 ~{\rm MeV}~=168.1~ {\rm MeV}~, \eqno (4.7)
$$
\noindent (particle symbols stand for their masses) and also
$$R=\frac{m_s-\hat{m}}{m_d-m_u} = \frac
{\Xi-\S}{n-p}=\frac{\S-N}{\Xi^{-}-\Xi^{0}}
$$
$$
60.7 \pm 9.1 = 46.2 \pm 5.9 .\eqno (4.8)
$$

\noindent In (4.8) the photon exchange contributions have been subtracted out
of the iso-spin violating mass differences.  The large deviation between
the central values in (4.8) shows that the  values of $R$ indicated there may
not be reliable.  Together with the observed deviation in (4.7) this suggests
that we go on to second order.  There is a known difficulty$^{8,10}$ in going
beyond first order in the CPT scheme since non-analytic terms like
$m^{\frac{3}{2}}$ and $m\ln m$ arise.  Since these are due to massless
Goldstone
boson exchange it would seem that this effect is already taken into account at
zeroth order if we expand around the point (4.6).  Recently, a new relation at
second order was found$^{25}$:
$$
R=\frac{3\L+\S-2N-2\Xi}{2\sqrt{3}m_{T}+(n-p)+(\Xi^{0}-\Xi^{-})}, \eqno (4.9)
$$
\noindent where $m_T$ is the ``corrected'' value of the $\L-\S^0$ transition
mass.  In principle, $m_T$ may be determined from precision measurements of the
difference between, say, $pK^-\rt\L\h$ and $n\bar{K}^0\rt \L\h$.  Thus (4.9)
may
be predictive in the future.  For the moment the best we can do is to use the
structure of (4.9) and the assumption that second order quantities don't
deviate by more than 20\% from their first order values to derive a lower
bound,
$$R>38\pm 10.\eqno (4.10)
$$

\noindent However this isn't very precise and we are forced to end our detour
by concluding that the quark mass ratios are not known well enough from
``outside'' to be useful for our purpose.  The fundamental quark mass ratios
should, in fact, thus be obtained as results from our fit.

The formulas for the meson one, two and three point functions used for fitting
are fairly standard and are explicitly given in ref. 1. It is useful to keep
$x=m_s/\hat{m}$ as a parameter during the fit and to evaluate everything for
each value of x. Then once the other decay widths and masses are used as
inputs we have four remaining predictions$^1$ for each x. These are displayed
below with the experimental values indicated in parentheses:
$$
\begin{array}{lllll}
x & (K^{*0}-K^{*+})_{\rm{NON-EM}} & \frac{\G(\r\rt 2\p)}
{\G(K^{*}\rt K\p)} & \frac{\G(\r \rt 2\p )}{\G(\f \rt K \bar{K})} & m(\f)\\
\hline
20 & 2.44~{\rm MeV} & 4.97 & 163 & 1.02~ {\rm GeV} ~(1.02)\\
25 & 3.05 & 4.45 & 81 & 1.04 \\
30 & 3.97 & 3.90 & 57 & 1.09 \\
32 & 4.47 & 3.67 & 47 & 1.13 \\
36 & 5.84 \left( \begin{array}{c}
5.2\\
7.4
\end{array}
\right)
 & 3.21   \begin{array}{c}
\phantom{aa} \\
(3.0)
\end{array}
 & 28  \begin{array}{c}
(40.3)\\
\phantom{aa}
\end{array}
& 1.26
\end{array}
$$

\noindent (Note that the experimental value for the non-electromagnetic part of
the $K^{*0}-K^{*+}$ mass difference is $5.2$ MeV if the individual masses given
in the Particle Data Tables are simply subtracted  but is 7.4 MeV if attention
is restricted to the ``dedicated'' experiments.)  We see that three quantities
are well fit for $x$ in the range $34-38$ but that $m(\f)$ is fit for $x=20$.
A compromise ``best fit'' is characterized by the fundamental quark mass
ratios:
$$
x=31.5,\sp y=-0.42,\sp R=36. \eqno(4.11)
$$

\noindent This may be compared with the values obtained by Gasser and Leutwyler
in their big Physics Report$^{10}$:

$$x=25.0 \pm 2.5, \sp y= -0.28 \pm 0.03, \sp R=43.5 \pm 2.2. \eqno(4.12)
$$

\noindent As a kind of check we note that we would also get y= -0.27 when we
set $x=25$.  Note that a recent determination$^{26}$ using $\j$ decays finds
$y=-0.54 \pm 0.09$ which is much closer to our result than to (4.12).

To see what physics lies behind this ``numerology'' let us attempt to
understand why our result (4.11) differs from (4.12).  We remark that the
Gasser-Leutwyler values are essentially what one gets by applying first order
perturbation theory to the vector multiplet.  However, there are some problems
with this.  For one thing the SU(3) relation between
$(M_{\r\o})_{\rm{NON-EM}}$, the quark mass induced part of the $\r\o$
transition mass, and $(K^{*0}-K^{*+})_{\rm{non-EM}}$ appears to be very badly
broken.  In fact, they used$^{10}~(M_{\r\o})_{\rm {non-EM}}$ as an input and
assumed something was wrong with the extraction of
$(K^{*0}-K^{*+})_{\rm{non-EM}}$ from experiment.  Furthermore even
though the width relation between
$\G(K^*\rt K\p)$ and
$\G(\r \rt \p\p)$ was one of the historical successes of SU(3) invariance, the
measured width of the $\r$ has increased from 100 MeV to 150 MeV since those
days and the simple SU(3) relation no longer holds.
Also the ordinary SU(3) width relation between $\G(\f\rt K\bar{K})$ and
$\G(\r \rt 2 \p)$ does not hold.  In the present approach $(K^{*0}-K^{*+}),
{}~~\G(K^*\rt K\p) / \G(\r \rt 2\p)$ and $\G(\f\rt K \bar{K})/
\G(\r \rt 2\p)$ can be explained with the help of a relatively important
derivative symmetry  breaker for the vectors (the $\g^\prime$ term in (4.5)).
This modifies the SU(3) relations via non-trivial wave-function
renormalizations for vector mesons containing strange quarks; explicitly$^1$,
$$
Z_\r=0.99,\sp Z_{K^{*}}=0.84, \sp Z_{\f}=0.65 \eqno(4.13)
$$ for the best fit parameters (Note, the physical $\r$ field, $\r_p$, is given
by
$\r_p=Z_\r\r ~\rm{etc.}~)$.  To fine tune our prediction for $m(\f)$, more
exotic terms and loop corrections should be included.  In any case it is clear
that the above feature represents a new qualitative effect for describing the
vector nonet system.

For our purposes the $\h^\prime$ -OZI rule violating-interactions are also very
important.  To mock up the axial anomaly and produce an $\h^\prime$ mass we add
the terms:

$$\frac {1}{\k} G^2 +\frac {i}{12} G~\ln (\det U/ \det U^\dagger) + n
{}~{\rm Tr}~ (\a_\m)~{\rm Tr}~(\a_\m)+ i\e^\prime G~{\rm
Tr}~[\cm(U-U^\dagger)], \eqno (4.14)$$

\noindent where $\a_\m \equiv \pa_\m UU^\dagger$ and $G=\pa_\m K_\m$.
No kinetic energy terms are included for $G$ so it acts as an auxiliary field
and gets eliminated by its equation of motion:
$$G=\frac {\sqrt{3} \k}{8} F_\p\h^\prime -
\frac{i\k \e^\prime}{2} ~{\rm Tr}~[\cm(U-U^{\dagger })], \eqno (4.15)
$$
\noindent which is to be substituted back into (4.14).  The Lagrangian which
results$^{27}$  when one keeps just the first two terms in (4.14) (i.e. setting
$n=\e^\prime=0)$ has been known for quite a while to be able to explain the
$\h'$ mass and the $\h-\h^\prime$ mixing angle but not $^{28}$ the $\h$ mass.
With the two additional
terms$^{29}$ we are now able to also explain the $\h$ mass as well as $\p^0\rt
\g\g,~\h\rt \g\g$ and $\h^\prime \rt \g\g$.  Note that substituting (4.15) into
(4.14) yields, among other things, a symmetry breaker of ``type 7'' but not one
of ``type 6'' in the second order CPT program$^8$.  Thus the assumption that
the
OZI rule violating interactions are dominated by the psudoscalar iso-singlet
channel (and given by (4.14)) evidently selects a unique ``frame'' from the one
parameter family allowed by the Kaplan-Manohar ambiguity$^{30}$.

With the present Lagrangian explaining more features of low energy meson
physics, we may feel more confident about investigating symmetry breaking in
the soliton sector.  This aspect is discussed in refs. 1 and 9.
Qualitatively, the effect of the important $\g^\prime$ term in (4.5) is to
decrease the relative contribution of the $\h$ excitation
$(\propto \d^\prime)$ to the
$n-p$  mass difference while leaving the net prediction about the same.
This has the consequence$^1$ that, while the ``glue'' contribution to $H(0)$
remains relatively small, it could be somewhat larger than the value in (3.17).

The question of symmetry breaking in the pseudoscalar-vector system is
certainly of interest outside the soliton sector too.  It seems interesting to
also calculate loop diagrams and include other ``second-order'' interaction
terms (although we believe we have the dominant symmetry breakers) to construct
an analog of the CPT scheme which would be useful up to about 1 GeV.  Of
course, this covers a lot of physics so we expect that progress in this
direction will be incremental or evolutionary in character.
\vglue 1.0cm
{\bf\noindent Acknowledgements}
\vglue 0.5cm
I would like to thank the organizers for holding an exceptionally pleasant and
worthwhile workshop.

This work was supported in part by the U.S. Department of Energy under contract
number DE-FG-02-85ER40231.
\vglue 1.0cm
{\bf\noindent References}
\vglue 0.5cm

\begin{enumerate}

\item J. Schechter, A. Subbaraman and H.
Weigel, Syracuse-T\"ubingen report, in preparation.

\item P. Jain, R. Johnson, N.W. Park, J. Schechter and H. Weigel,  {\it Phys.
Rev.}   {\bf D40} 855 (1989).

\item R. Johnson, N.W.Park, J. Schechter, V. Soni and H. Weigel, {\it Phys.
Rev.} {\bf D42}, 2998 (1990).

\item J. Schechter, V. Soni, A. Subbaraman and H. Weigel, {\it Phys. Rev.
Lett.}
{\bf 65}, 2955 (1990).

\item Progress, discussed in the talks of G. Holzwarth and B. Moussallam at
this workshop, has been made on the problem of too large absolute baryon
masses.

\item V. Bernard, N. Kaiser and U.-G. Meissner, {\it Phys. Lett.} {\bf B237},
545 (1990).

\item Reviews are provided by R. Jaffe and A. Manohar, {\it Nucl. Phys.}
{\bf B337}, 509 (1990); G. Altarelli and W.J. Stirling, CERN report Th 5249/88;
H.-Y. Cheng, Taipei report IP-ASTP-01-91. A ``brief review'' of the material in
sec. 3 is given in J. Schechter, V. Soni, A. Subbaraman and H. Weigel, {\it
Mod. Phys. Lett. A} {\bf 7}, 1 (1992); E {\bf 7}, 1199 (1992).

\item References may be traced from J. Gasser and H. Leutwyler, {\it Nucl.
Phys.} {\bf B250}, 465 (1985).

\item Talk of H. Weigel at this workshop.

\item This is reviewed in J. Gasser and H. Leutwyler, {\it Phys. Rep.} {\bf
87},
77 (1982).

\item See (5.6) of P. Jain, R. Johnson, Ulf-G. Meissner, N.W. Park and J.
Schechter, {\it Phys. Rev.} {\bf D 37}, 3252 (1988) and also Ulf-G. Meissner,
N. Kaiser, H. Weigel and J. Schechter, {\it Phys. Rev.} {\bf D39}, 1956 (1989).

\item H. Yabu and K. Ando, {\it Nucl. Phys.} {\bf B301}, 601 (1988).

\item H. Weigel, J. Schechter, N.W. Park and Ulf-G. Meissner, {\it Phys.
Rev.} {\bf D42}, 3177 (1990).

\item E.M. Collaboration, J. Ashman et al., {\it Phys. Lett.} {\bf B206}, 364
(1988); {\it Nucl. Phys.} {\bf B328}, 1 (1989).

\item N.W. Park, J. Schechter and H. Weigel, {\it Phys. Lett.} {\bf B228}, 420
(1989); {\it Phys. Rev.} {\bf D41}, 2836 (1990); {\it Phys. Rev.} {\bf D43},
869 (1991); N.W. Park and H. Weigel, {\it Nucl. Phys.} {\bf A541}, 453 (1992).
The possibility of such a situation in the quark model approach is discussed by
H. Lipkin, {\it Phys. Lett.} {\bf B256}, 284 (1991).

\item S. Brodsky, J. Ellis and M. Karliner, {\it Phys. Lett.} {\bf B206}, 309
(1988).

\item See the first of ref. 15 and T. Cohen and M. Banerjee, {\it Phys. Lett.}
{\bf B230}, 129 (1989).

\item It is explicitly given in (4.5) of ref. 3.

\item R. Carlitz, J. Collins and A. Mueller, {\it Phys. Lett.} {\bf B214}, 229
(1988); A.V. Efremov and D.J. Teryaev, Dubna report JINR E2-88-297; G.
Altarelli and G.G. Ross, {\it Phys. Lett. } {\bf B212}, 391 (1989).

\item In addition to the reviews in ref. 7, see J. Mandula, {\it Phys. Rev.
Lett.} {\bf 65}, 1403 (1990).

\item G.M. Shore and G. Veneziano, {\it Phys. Lett. } {\bf B244}, 75 (1990).
See also J. Schechter, V. Soni, A. Subbaraman and H. Weigel, {\it Mod. Phys.
Lett.} {\bf A5}, 2543 (1990); T. Hatsuda, Univ. of Washington report INT-5-90.

\item See A. Efremov, J. Soffer and N. Tornqvist, {\it Phys. Rev. Lett.} {\bf
64}, 1495 (1990).

\item See ref. 11 and references therein.

\item S. Okubo, {\it Phys. Lett.} {\bf 5}, 165 (1963).

\item J. Schechter and A. Subbaraman, {\it Int. Jour. Mod. Phys. A} to be
published.

\item  J. Donoghue and D. Wyler, {\it Phys. Rev.  } {\bf D45}, 892 (1992).

\item C. Rosenzweig, J. Schechter and G. Trahern, {\it Phys. Rev.} {\bf D21},
3388 (1980); P. Di Vecchia and G. Veneziano, {\it Nucl. Phys. } {\bf 253}
(1980); E. Witten, {\it Ann. Phys.} {\bf 128}, 1789 (1981); P. Nath and R.
Arnowitt, {\it Phys. Rev.} {\bf D23}, 473 (1981); K. Kawarabayashi and N. Ohta,
{\it Nucl. Phys.} {\bf B175}, 477 (1980); A. Aurilia, Y. Takahashi and P.
Townsend, {\it Phys. Lett.} {\bf B95}, 265 (1980).

\item V. Mirelli and J. Schechter, {\it Phys. Rev. } {\bf D15}, 1361 (1977).

\item See also P. Di Vecchia, F. Nicodemi, R. Pettorino and G. Veneziano, {\it
Nucl. Phys.} {\bf B181}, 318 (1981).

\item D. Kaplan and A. Manohar, {\it Phys. Rev. Lett.} {\bf 56}, 2004 (1986).

\end{enumerate}
\end{document}